\documentclass[preprint,12pt, showkeys]{revtex4}

\usepackage{color}
\usepackage[brazil, english]{babel}
\usepackage[utf8]{inputenc}
\usepackage{graphicx}
\usepackage{epsfig}
\usepackage{amssymb}
\usepackage{amsmath} 
\usepackage{slashed}
\usepackage[unicode=true,bookmarks=false,breaklinks=false,pdfborder={0 0 1},colorlinks=true]
 {hyperref}
\hypersetup{
 citecolor=blue,linkcolor=blue,urlcolor=blue}

\graphicspath{%
    {converted_graphics/}
    {/}
}
\begin{document}

\title{Proton Structure Functions from an AdS/QCD model with a deformed background} 
\author{Eduardo Folco Capossoli$^{1,2}$}
\email{eduardo\_capossoli@cp2.g12.br} 
\author{Miguel Angel Martín Contreras$^3$}
\email{miguelangel.martin@uv.cl}
\author{Danning Li$^4$}
\email{lidanning@jnu.edu.cn}
\author{Alfredo Vega$^3$}
\email{alfredo.vega@uv.cl} 
\author{Henrique Boschi-Filho$^{1}$}
\email{boschi@if.ufrj.br}  
\affiliation{$^1$Instituto de F\'\i sica, Universidade Federal do Rio de Janeiro, 21.941-972 - Rio de Janeiro-RJ - Brazil \\
 $^2$Departamento de F\'\i sica / Mestrado Profissional em Práticas de Educação Básica (MPPEB), Col\'egio Pedro II, 20.921-903 - Rio de Janeiro-RJ - Brazil\\
 $^3$Instituto de Física y Astronomía, Universidad de Valpara\'iso, A. Gran Breta\~na 1111, Valpara\'iso, Chile\\ 
$^4$Department of Physics and Siyuan Laboratory, Jinan University, Guangzhou 510632, China}

\begin{abstract}
In this work we study unpolarized spin $1/2$ baryonic deep inelastic scattering (DIS) in the regime of large Bjorken parameter $x$. We calculate the corresponding structure functions $F_{1,2}(x,q^2)$. Our approach is based on an AdS/QCD model with a deformed background, where we consider an exponential factor in the AdS$_5$ metric. Such a deformation implies the introduction of an anomalous dimension in the model.  Our results for the structure functions are consistent with those found in the literature from experimental data. 
\end{abstract}


\maketitle


\section{Introduction}

Since the beginning of the study of elementary particle under high energy processes the usage of the Deep inelastic scattering (DIS) became it as one of the most iconic experiment in the this field, allowing to probe the proton and neutron internal structures. The canonical way to study DIS is based on a non-abelian gauge field theory known as QCD.  
In its fundamental works   \cite{Georgi:1951sr, Gross:1974cs, Gross:1976xt, Jaroszewicz:1982gr}, it has been shown that the canonical dimension $\Delta_{\rm can}$ of an operator $\cal O$ should be modified by the introduction of an anomalous contribution $\gamma$, which implies that $[{\cal O}]=\Delta_{\rm can} + \gamma$. 
In particular, in Refs. \cite{Georgi:1951sr, Gross:1976xt, Jaroszewicz:1982gr} the authors addressed the anomalous dimension in the context of DIS. Note however the QCD perturbative techniques are not reliable at low energies. 

An alternative way to deal with this problem is based on the anti-de Sitter/Conformal Field Theory (AdS/CFT) correspondence  which relates a conformal ${\cal N} = 4$ super Yang-Mills theory (SYM) with symmetry group $SU(N)$, for $N \rightarrow \infty$, living in $3+1$ dimensional Minkowski spacetime to a superstring theory in a 10-dimensional curved spacetime (for a review see for instance \cite{Aharony:1999ti}).

After breaking conformal invariance one has a phenomenological holographic approach known as AdS/QCD. In this context many important works dealt with DIS providing interest results as can be seen for instance in Refs. \cite{Polchinski:2002jw, Hatta:2007he, BallonBayona:2007rs, BallonBayona:2007qr, Cornalba:2008sp, Pire:2008zf, Albacete:2008ze, BallonBayona:2008zi, Gao:2009ze, Taliotis:2009ne, Yoshida:2009dw, Hatta:2009ra, Avsar:2009xf, Cornalba:2009ax, Bayona:2009qe, Cornalba:2010vk, Brower:2010wf, Gao:2010qk, BallonBayona:2010ae, Braga:2011wa, Koile:2013hba, Koile:2014vca, Gao:2014nwa, Capossoli:2015sfa, Koile:2015qsa, Jorrin:2016rbx, Kovensky:2016ryy, Kovensky:2017oqs, Amorim:2018yod, deTeramond:2018ecg, Liu:2019vsn,  Watanabe:2019zny, Jorrin:2020cil}. Most of these references studied DIS for scalar particles and some of them for vector mesons and baryons. In particular, the pioneer work presented in Ref. \cite{Polchinski:2002jw}, treated holographic DIS within hardwall model for scalars and fermions taking into account the regimes of large, small and exponentially small for the Bjorken parameter $x$.  In refs. \cite{BallonBayona:2010ae, Koile:2013hba} the authors studied DIS for vector particles, while in Refs. \cite{BallonBayona:2007qr, Gao:2009ze, Gao:2010qk, Braga:2011wa, Jorrin:2020cil} for baryonic DIS.

Regarding the anomalous dimension, many works dealt with it within the holographic context,   for instance, \cite{Gubser:2002tv, Armoni:2006ux,  Brower:2006ea,  Gubser:2008yx, Vega:2008te, Braga:2014pxa, FolcoCapossoli:2016ejd, BoschiFilho:2012xr, Capossoli:2016ydo}. 
In particular, in Ref. \cite{Brower:2006ea} the authors have argued about a possible introduction of an anomalous dimension  for DGLAP and BFKL regimes. In Ref. \cite{Vega:2008te} the authors used such anomalous dimension to fit the masses for mesons and baryons. In Ref. \cite{Braga:2011wa}, the authors considered anomalous dimensions in the holographic description of DIS. 
In Refs. \cite{BoschiFilho:2012xr, Capossoli:2016ydo} the authors have used the anomalous dimension related to the QCD beta function in order to compute even and odd glueball masses.  Interestingly, in Refs. \cite{Gubser:2008yx, FolcoCapossoli:2016ejd} the anomalous dimension decreased the value of the  conformal dimension, without violating the Breitenlohner-Freedman bound. 

Here in this work we will use a holographic model to compute the baryonic DIS structure functions, $F_1 = F_1 (x, q^2)$ and  $F_2 = F_2 (x, q^2)$ for the proton, which are  dependent on the photon virtuality $q^2$, and the Bjorken parameter $x$. In particular our focus is on the large $x$ regime. To perform our computation we will use an  AdS/QCD model with a deformed $AdS_5$ background. This deformation of the $AdS_5$ space breaks the conformal invariance and generates a mass scale for the fermionic fields. Such a deformation was considered before in a wide range of AdS/QCD studies, as can be seen in Refs. \cite{Andreev:2006vy, Andreev:2006ct, Ghoroku:2005vt, Forkel:2007cm, White:2007tu, Wang:2009wx, Afonin:2012jn, Rinaldi:2017wdn, Bruni:2018dqm, Diles:2018wbe, Afonin:2018era,  FolcoCapossoli:2019imm, Rinaldi:2020ssz, Tahery:2020tub, Caldeira:2020sot}.
Our DIS structure functions are compared with experimental data showing good agreement for large $x$. 

This work is organized as follows: in section \ref{secdis} we review briefly the main properties of DIS.  In section \ref{defdis} we present our deformed AdS space model which describes the interaction between a vector and spinor fields. In particular we compute holographically the wave functions for these  fields. In section \ref{disintaction} we compute the DIS interaction action and extract the expressions for the structure functions $F_1(q^2, x)$ and $F_2(q^2, x)$. In Section \ref{numerical} we present our numerical analysis for the structure functions and compare them with available experimental data. In Section \ref{conc} we present our conclusion and discussions.

\section{Brief review of DIS}\label{secdis}
\begin{figure}[!ht]
  \centering
  \includegraphics[scale = 0.70]{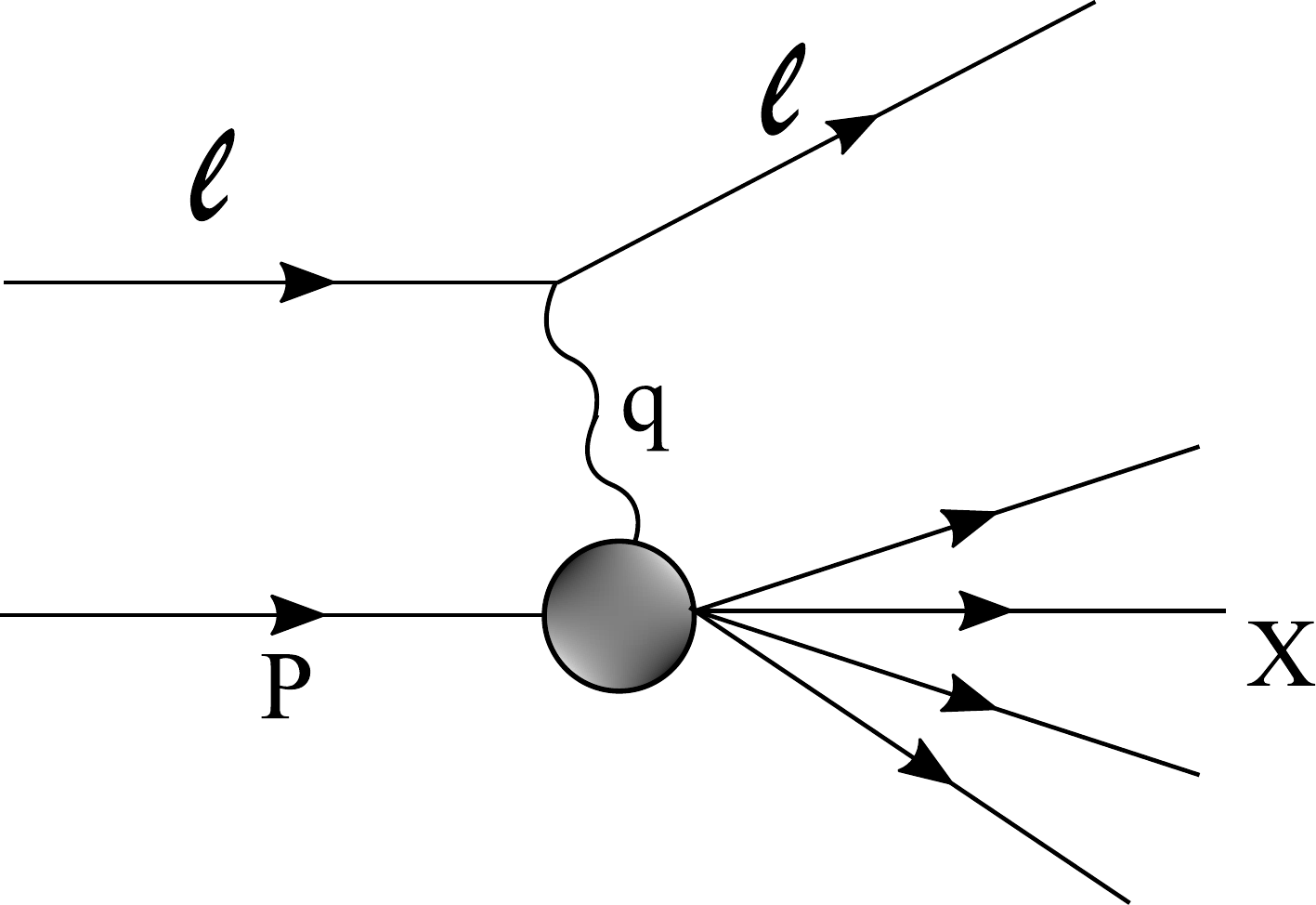} 
\caption{Deep inelastic scattering between a hadron and a lepton through the exchange of a virtual photon.}
\label{dis}
\end{figure}

Scattering processes play an essential role in particle physics since they allow us to explore most of the hadronic properties. In particular, \emph{deep inelastic scattering} (DIS) is the tool that probes the inner hadronic structure.  This process consists of a lepton scattered off a proton target, causing its fragmentation into other hadronic states. In figure \ref{dis}, we depict the Feynmann diagram for the process. It is the next most straightforward reaction involving strong interactions, after the $e^+e^-\to$ hadrons. 

In a schematic point of view, DIS is an electromagnetic scattering off a charged parton, i.e., a quark, inside the proton by the incident lepton, which can be an electron or a muon.  If the four-momentum transferred by the lepton to the proton target is large, the inner quark is expelled out from the target. In the process, the quark radiates gluons and quark-antiquark pairs that will hadronize soon after.


To consider the DIS process quantitatively, we will discuss the following reaction: $l\,p\to \,l\,X$, where the final hadronic state $X$ will label all of the produced hadrons by the proton fragmentation. We can determine from the fragmentation the inner structure of the target proton.  The so-called \emph{Bjorken variable} parametrizes this fragmentation according to: 

\begin{equation}
x = - \frac{q^2} {2 P \cdot q}\,, 
\end{equation}

\noindent where $q^2$ is the transferred momentum from the lepton to the proton by a virtual photon and $P$ is the initial proton momentum, with mass defined as $P^2=-M^2$.  After setting the kinematical frame, now we can write the scattering amplitude as:

\begin{equation}
i\,\mathcal{M}_{l\,p\to l\,X}=(i\,Q)\,\bar{u}\,\gamma_\mu\,u\,\left(\frac{i}{q^2}\right)\,(i\,e)\int{d^4y\,e^{i\,q\cdot y}\langle X\left|J^\mu (y)\right|P\rangle},    
\end{equation}

\noindent where $J_\mu(x)$ is the quark electromagnetic current. The crucial step in this analysis is how to connect the proton fragmentation with the emergence of highly energetic hadrons in the final state. This information is encoded into the hadronic tensor, constructed from the current between the proton and the X final state. Since the virtual photon is responsible for breaking the proton,  we can use the optical theorem 

\begin{equation}
\sum_X{\int{d\,\Pi_X\,\left|\mathcal{M}_{\gamma\,p\to X}\right|^2}}=2\mathbb{I}\text{m}\,\mathcal{M}_{\gamma\,p\to \gamma\,p}    
\end{equation}
to write the hadronic transition amplitude $W^{\mu\,\nu}$, in terms of the forward matrix element of two proton currents averaged over the spin:

\begin{equation}\label{f3}
W^{\mu \nu} = \frac{i}{4\, \pi} \sum_s \int d^4 y~ e^{i q.y} \langle P, s|\mathcal{T}\left\{J^{\mu}(y)\, J^{\nu}(0)\right\}|P, s \rangle.
\end{equation}
This expression for the hadronic tensor is known as \emph{the forward Compton scattering amplitude}. Notice that $|P, s \rangle$ represents a normalizable proton state with spin $s$,  $J^{\mu}$ is the electromagnetic quark current introduced before  (for a review see \cite{Manohar:1992tz}) and $\mathcal{T\{ {\cal O}_{\rm 1}{\cal O}_{\rm 2} \}}$ means that the operators product is temporally ordered.

For the imaginary part of this amplitude, we can write, in the Fourier space, as 

\begin{equation}\label{ImWmunu}
\mathbb{I}\text{m}\,W^{\mu\nu}=\frac{1}{4}\sum_X\,\delta\left(M_X^2-(P+q)^2\right)\langle P,s\left|J^\mu(0)\right|X\rangle\langle X\left|J^\nu(0)\right|P,s\rangle,    
\end{equation}

\noindent where the final state $\left|X\right.\rangle$ is characterized by the total invariant mass $M_X$, constrained by the energy conservation, i.e.,  $M_X>M$.  

To go further into this analysis, we have to address the form of the hadronic tensor. To do so, the Ward-Takahashi identity requires that  
\begin{equation}
q_\mu\,W^{\mu\nu}=q_\nu\,W^{\mu\nu}=0.     
\end{equation}

This condition allows us to write the hadronic tensor as a decomposition into scalar functions $F_{1,2}\equiv F_{1,2}(x,q^2)$, defined in terms of invariant quantities, as follows:

\begin{equation}\label{f4}
W^{\mu \nu} = F_{1} \left( \eta^{\mu \nu} - \frac{q^{\mu} q^{\nu}}{q^2} \right) + \frac{2x}{q^2} F_2 \left( P^{\mu} + \frac{q^{\mu}}{2x} \right)\left( P^{\nu} + \frac{q^{\nu}}{2x} \right).
\end{equation}
We have just written the symmetric terms that will remain after the contraction with the leptonic tensor, which is symmetric. Notice also that for spin-1/2 targets, the hadronic tensor decomposed as in Eq. \eqref{f4} is independent on the final hadron spin. The spin dependence is encoded into the antisymmetric terms of $W^{\mu\nu}$, relevant only for the non-physical region $x>1$.  This particularity makes the hadronic DIS cross-section dependent on combinations that have both hadron and lepton spins or none of them. For our purposes, we are going to consider only unpolarized leptons and target protons.


\section{The deformed A\lowercase {d}S space model and the DIS}\label{defdis}

 We start this section discussing the deformed AdS/QCD model which will be used to calculate the DIS structure functions for fermionic targets. The action for the fields can be written as: 
\begin{equation}\label{acao_soft}
S = \int d^{5} x \sqrt{-g} \; {\cal L}
\end{equation}

\noindent where ${\cal L}$ is the Lagrangean density, $g$ is the determinant of the metric $g_{mn}$ of the deformed $AdS_5$ space, given by:
\begin{equation}\label{gs}
ds^2 = g_{mn} dx^m dx^n= e^{2A(z)} \, (dz^2 + \eta_{\mu \nu}dy^\mu dy^\nu)\,. 
\end{equation}

\noindent Here we have considered the AdS radius $R=1$, $z$ is the holographic coordinate and 
\begin{equation}
    A(z) = -\log z + \frac{k}{2}\, z^2\,. \label{A}
\end{equation}
\noindent 
The constant $k$ has dimension of mass  squared and is associated with a QCD mass scale. 
In this work we use indices $
m, n, \cdots$  to refer to the $5-$dimensional space, separating into $\mu, \nu, \cdots$ for the Minkowski spacetime and the holographic $z$ coordinate. The coordinates $x^\mu$ have signature $(-,+,+,+)$ and also describe the boundary of the deformed AdS space where the gauge theory lives. 

Note that the  metric given by Eqs.  \eqref{gs} and \eqref{A}  represent a deformed AdS space since we introduced the warp factor $e^{k z^2}$ into its definition.  The present model is inspired by Refs. \cite{Andreev:2006vy, Andreev:2006ct}
where this warp factor was introduced in the AdS metric to obtain the quark-antiquark potential. This model was used recently to obtain the hadronic spectrum of particles with various spins including spin 1/2 fermions \cite{FolcoCapossoli:2019imm}, which is relevant to the present discussion of DIS with baryonic target (see our discussion on subsection \ref{barstates}). 

This deformed AdS background formulation can be compared with the original softwall model \cite{Karch:2006pv}. Actually, they produce different equations of motion despite that both imply linear confinement. In the particular case of the fermionic sector, the dilaton in the action of the softwall does not couple to bulk fermions, meaning that one can not get a discrete spectrum.

In order to get a discrete spectrum for the fermionic sector, one needs to modify the softwall model introducing a hardwall as in Ref. \cite{BallonBayona:2007qr}, producing a hybrid model, or introducing a $z$ coordinate dependent mass term as in Refs.\cite{Braga:2011wa,Gutsche:2011vb}. On the other hand, in our deformed background model, the fermionic discrete spectrum emerges naturally due to the geometry  of the AdS space modified by the introduction of a quadratic exponential warp factor.

At this point let us briefly discuss the holographic approach to DIS, inspired by Ref.  \cite{Polchinski:2002jw} in the supergravity approximation for string theory in the large $x$ regime.

Following the holographic dictionary we will connect the matrix element of canonical DIS given by Eq.\eqref{f4} with the supergravity interaction action in AdS space, $S_{\rm int}$. Considering that the baryonic particle was scattered off by a virtual photon with polarization $\eta_{\mu}$, one can write: 
\begin{equation}\label{corrint}
   \eta_\mu \langle P + q, s_X | J^{\mu}(0)|P, s\rangle = S_{\rm int}
\end{equation}

\noindent where the interaction action is given by:
\begin{eqnarray}\label{sintdis}
S_{\rm int}&=& g_V\,\int{dz\,d^4y\,\sqrt{-g}\,\phi^\mu\,\bar{\Psi}_X\,\Gamma_\mu\,\Psi_i}\,,
\end{eqnarray}
\noindent with $g_V$ a coupling constant related to the electric charge of the baryon and $\Gamma_{\mu}$ are Dirac gamma matrices in curved space. 
The spinors $\Psi_i$ and $\Psi_X$ are the initial and final states for the baryon and $\phi^{\mu}$ is the electromagnetic gauge field. All those quantities will computed in the following sections.

\subsection{Computing the electromagnetic field}

Since DIS also involves an electromagnetic interaction, in this section we will describe the photon in the deformed AdS space.

Let us to introduce the action for a five dimensional massless gauge field $\phi^n$ given by:
\begin{equation}\label{f15}
S = -  \int d^{5} x \sqrt{-g} \; \frac{1}{4} F^{mn} F_{mn}\,,
\end{equation}
\noindent where $F^{mn} = \partial^m \phi^n - \partial^n \phi^m$. This action leads to the following equations of motion
\begin{equation}\label{f16}
\partial_m [ \sqrt{-g}\; F^{mn}] = 0\,. 
\end{equation}

Using the gauge fixing 
\begin{equation}
\partial_\mu\,\phi^\mu+e^{-A}\partial_z\left(e^A\,\phi_z\right)=0,  
\end{equation}
where $A=A(z)$ is given by Eq. \eqref{A},   one has 
\begin{eqnarray} \label{solem}
\Box\,\phi_\mu+A'\,\partial_z\,\phi_\mu+\partial_z^2\,\phi_\mu=0\\
\Box\,\phi_z-\partial_z\left(\partial_\mu\,\phi^\mu\right)=0\,, \label{solem2}
\end{eqnarray}
\noindent where prime denotes derivative with respect to $z$. 

Just before we present the solutions for Eqs. \eqref{solem}  and \eqref{solem2} it is worthy to mention that we will consider, for a sake of simplicity, and without loss of generality, a photon with a particular polarization such that $\eta_ \mu \, q^ \mu =0$. In this sense only  the electromagnetic field component $\phi^\mu$ will contribute in the scattering process as discussed, {\sl e. g.}, in Refs. \cite{Polchinski:2002jw, BallonBayona:2007qr, Braga:2011wa}.

The general solution to equation \eqref{solem} has the following form: 
\begin{eqnarray}
\phi_\mu(z,q)=C^1_{\mu}(y)\, G_{1,2}^{2,0}\left(\frac{k\,z^2}{2}\left|
\begin{array}{c}
 \frac{q^2}{2 k}+1 \\
 0,1 \\
\end{array}
\right.\right)-\frac{1}{2} C^2_{\mu}(y)\, k\, z^2 \, _1F_1\left(1-\frac{q^2}{2 k};\,2;\,-\frac{k\,z^2}{2}\right)\,,
\end{eqnarray}
\noindent where 
\begin{equation}
    G_{p,q}^{m,n}\left(z\left|
\begin{array}{c}
 a_1 \cdots a_p \\
 b_1 \cdots b_q \\
\end{array}
\right.\right)\,\,\, {\rm and}\,\,\, _1F_1 (a; b; z)\nonumber 
\end{equation}

\noindent are the the Meijer G function and the Kummer confluent hypergeometric function, respectively. By imposing the boundary condition $\left.\phi_{\mu}(z,y)\right|_{z=0} = \eta_{\mu} e^{i q\cdot y}$, that implies $C_\mu^1(y)=0$, and considering normalizable (square integrable) solutions, one can write:
\begin{eqnarray}\label{phimunorm}
\phi_\mu(z,q)&=&-\frac{\eta_{\mu} e^{i q\cdot y}}{2} \, k\, z^2 \, \Gamma{\left[1 - \frac{q^2}{2k}\right]}\; {\cal U} \left(1-\frac{q^2}{2 k};\,2;\,-\frac{k\,z^2}{2}\right)\nonumber \\
&\equiv& -\frac{\eta_{\mu} e^{i q\cdot y}}{2}\, B(z,q)\,, 
\end{eqnarray}
\noindent where $\Gamma[a]$ is the Gamma function and ${\cal U} (a, b,z)$ is the Tricomi hypergeometric function  \cite{abramowitz}. 
This equation represents the solution for the electromagnetic field that will be used to compute interaction action in Eq. \eqref{sintdis}.

\subsection{Computing the baryonic states} \label{barstates}

In order to obtain the interaction action $S_{\rm int}$, Eq. \eqref{sintdis}, one needs to compute the initial and final baryonic states. 
The action for the fermionic fields in the deformed AdS space can be written as  
\begin{equation}\label{diracfield}
S =  \int d^{5} x\sqrt{g} \; \bar{\Psi}({\slashed D} - m_5 ) \Psi, 
\end{equation}
with the operator ${\slashed D}$ defined as:
\begin{equation}\label{slash}
{\slashed D} \equiv g^{mn} e^{a}_n \gamma_a \left( \partial_{m} + \frac{1}{2} \omega^{bc}_{m}\, \Sigma_{bc} \right) = e^{-A(z)} \gamma^5 \partial_5 + e^{-A(z)} \gamma^{\mu}  \partial_{\mu} + 2 A'(z)\gamma^5, 
\end{equation}
\noindent where $\gamma_a = (\gamma_{\mu}, \gamma_5)$, $\left\lbrace \gamma_a, \gamma_b \right\rbrace = 2 \eta_{ab} $, and $\Sigma_{\mu 5} = \frac{1}{4} \left[ \gamma_{\mu}, \gamma_5\right]$. This prescription follows from the pure AdS space given in Refs. \cite{Henningson:1998cd, Mueck:1998iz, Kirsch:2006he, Abidin:2009hr}.  
The Dirac's gamma matrices are represented by $\gamma_\mu$ and we will use use $a, b, c$ to represent indexes in flat space,  $m, n, p, q$ to represent  indexes in the deformed $AdS_5$ space, and $\mu, \nu$ to represent the Minkowski space.
Thus, the vielbein are given by:
\begin{equation}\label{tetra}
e^{a}_m = e^{A(z)} \delta^{a}_m, \; \;\; e^{m}_a = e^{-A(z)} \delta^m_{a} e^{ma} = e^{-A(z)} \eta^{ma}, \; \; \; {\rm with} \; \;\;\;\;m = 0, 1, 2, 3, 5.
\end{equation}

For the spin connection $\omega^{\mu \nu}_{m}$, one has:
\begin{equation}
\omega^{a b}_{m}  = e^a_n \partial_m e^{nb} + e^a_n e^{pb} \Gamma^n_{pm}, 
\end{equation}
where the Christoffel symbols are written as:
\begin{equation}\label{Cris}
\Gamma_{m n}^p = \frac{1}{2} g^{pq}(\partial_n g_{mq} + \partial_m g_{nq} - \partial_q g_{mn}), \;\;\; {\rm with} \;\;\;g_{mn} = e^{2 A(z)} \eta_{mn}.
\end{equation}
The only non-vanishing $\Gamma_{m n}^p $ for the deformed AdS space are:
\begin{equation}\label{LV}
\Gamma_{\mu \nu}^5 = A'(z) \eta_{\mu \nu}, \; \; \Gamma_{5 5}^5 = -A'(z)  \; \; {\rm and} \; \; \Gamma_{\nu 5}^{\mu} = -A'(z) \delta^{\mu}_{\nu},
\end{equation}
\noindent so that 
\begin{equation}\label{spin}
\omega^{5 \nu}_{\mu} = - \omega^{\nu 5} _{\mu} = \partial_z A(z) \delta^{\nu}_{\mu} 
\end{equation}
\noindent and all other components of the spin connection vanish.

From the action Eq. \eqref{diracfield} one can derive the EOM: 
\begin{equation}\label{fermioneom}
({\slashed D} - m_5 ) \Psi  = 0, 
\end{equation}
which can be written as:
\begin{equation}\label{newdirac}
\left( e^{-A(z)} \gamma^5 \partial_5 + e^{-A(z)} \gamma^{\mu}  \partial_{\mu} + 2 A'(z)\gamma^5 - m_5\right) \Psi = 0, 
\end{equation}
\noindent where $\partial_5 \equiv \partial_z$, and $m_5$ is the baryon bulk mass. Assuming that the spinor $\Psi$ can be decomposed into right- and left-handed chiral components, one has:
\begin{equation}\label{psi}
\Psi(x^{\mu}, z) = \left[ \frac{1 - \gamma^5}{2} f_L(z) + \frac{1 + \gamma^5}{2} f_R(z)\right] \Psi_{(4)}(x)\,, 
\end{equation}
\noindent where $\Psi_{(4)}(x)$ satisfies the usual Dirac equation $({\slashed \partial} - M)\Psi_{(4)}(x) = 0$ on the flat four-dimensional boundary space. For the left and right modes, one has $\gamma^5 f_{L/R} = \mp f_{L/R}$ and $\gamma^{\mu}  \partial_{\mu} f_R = M f_L$, and $M$ is the four-dimensional fermionic mass. 

Considering that the Kaluza-Klein modes are dual to the chirality
spinors one can expand:
\begin{equation}\label{kkmodes}
\Psi_{L/R} (x^{\mu}, z) = \sum_n f_{L/R}^n  (x^{\mu}) \chi_{L/R}^n (z).
\end{equation}

By using \eqref{kkmodes} with \eqref{psi} in \eqref{newdirac} one gets the coupled equations:
\begin{equation}\label{mix1}
\left(\partial_z + 2 A'(z)\, e^{A(z)} + m_5\,e^{A(z)} \right) \chi_{L}^n (z) = + M_n \chi_{R}^n (z)
\end{equation}
\noindent and
\begin{equation}\label{mix2}
\left(\partial_z + 2 A'(z)\, e^{A(z)} - m_5\,e^{A(z)} \right) \chi_{R}^n (z) = - M_n \chi_{L}^n (z).
\end{equation}

Performing a Bogoliubov transformation 
\begin{equation}\label{cv}
\chi^n_{L/R}(z) = \psi^n(z) e^{-2\, A(z)},
\end{equation}
\noindent and decoupling Eqs. \eqref{mix1} and \eqref{mix2}, one gets a Schr\"odinger equation written for both right and left sectors, given by:
\begin{eqnarray}\label{scr}
-\psi_{R/L}''(z) + \left[ m_5^2 e^{2 A(z)} \pm m_5 e^{A(z)}A'(z) \right]\psi_{R/L}(z)  = M_n^2 \psi_{R/L}^n (z),
\end{eqnarray}
\noindent where $M_n$ is the four-dimensional baryon mass for each mode $\psi^n_{R/L}$ and the corresponding potentials are given by:
\begin{equation}\label{potscr}
 V_{R/L}(z) = m_5^2 e^{2 A(z)} \pm m_5 e^{A(z)}A'(z). 
\end{equation}
One should note that this equation can be applied to any warp factor $A(z)$. The pure AdS space is recovered if one uses $A(z) = - \log(z)$, which leads to analytical solutions. In our case, with $A(z)= - \log z + kz^2/2$, Eq. \eqref{A}, we need to resort to numerical methods. 

From the solutions of Eq. \eqref{scr} one can read the final spinor state $\Psi_X$ and the initial spinor state  $\Psi_i$ as linear combinations of the chiral solutions $\psi_{R/L}$, as follows:
\begin{eqnarray}
\Psi_i&=&e^{i\,P\cdot\,y}\,z^2\,e^{-k\,z^2}\left[\left(\frac{1+\gamma_5}{2}\right)\psi^i_L(z)+\left(\frac{1-\gamma_5}{2}\right)\,\psi_R^i(z)\right]\,u_{s_i}(P)   \label{psii} \\
\Psi_X&=&e^{i\,P_X\cdot\,y}\,z^2\,e^{-k\,z^2}\left[\left(\frac{1+\gamma_5}{2}\right)\psi^X_L(z)+\left(\frac{1-\gamma_5}{2}\right)\,\psi^X_R(z)\right]\,u_{s_X}(P_X),  \label{psix}
\end{eqnarray}
where $s_i$ and $s_X$ are the spin of the initial and final states, respectively.

Let us comment here about the relation between $m_5$ and the conformal dimension $\Delta$. In pure AdS space, the bulk mass $m_5^{\rm AdS}$, according the AdS/CFT dictionary, is related to the canonical conformal dimension ($\Delta_{\rm can}$) of a boundary operator ${\cal O}$ as 
\begin{equation}
|m_5^{\rm AdS}|= \Delta_{\rm can} - 2. 
\end{equation}

 In its fundamental works   \cite{Georgi:1951sr, Gross:1974cs, Gross:1976xt, Jaroszewicz:1982gr}, it has been shown that the canonical dimension $\Delta_{\rm can}$ of an operator $\cal O$ should be modified by the introduction of an anomalous contribution $\gamma$, which implies an effective scaling dimension that $\Delta_{\rm eff}=\Delta_{\rm can} + \gamma$. Then $m_{5}$ could be changed by 
\begin{equation}\label{m5ano}
 |m_5|= \Delta_{\rm can} + \gamma - 2. 
\end{equation}
Hence, we will take into account this  anomalous dimension in our model to compute the structure functions in baryonic DIS, as described in Fig. \ref{dis}. We choose the initial state to be a proton, and for our purpose, it will be considered as a single particle (disregarding the internal constituents), as was done in Ref. \cite{Braga:2011wa}, with $\Delta_{\rm can} = 3/2$, which is the usual fermionic dimension. This could be justified looking at the proton's parton distribution functions (PDFs) as presented in PDG \cite{pdg}. In those PDG plots one can see that the PDFs go to zero for $x\to 1$.

\begin{figure}[!ht]
  \centering
  \includegraphics[scale = 0.70]{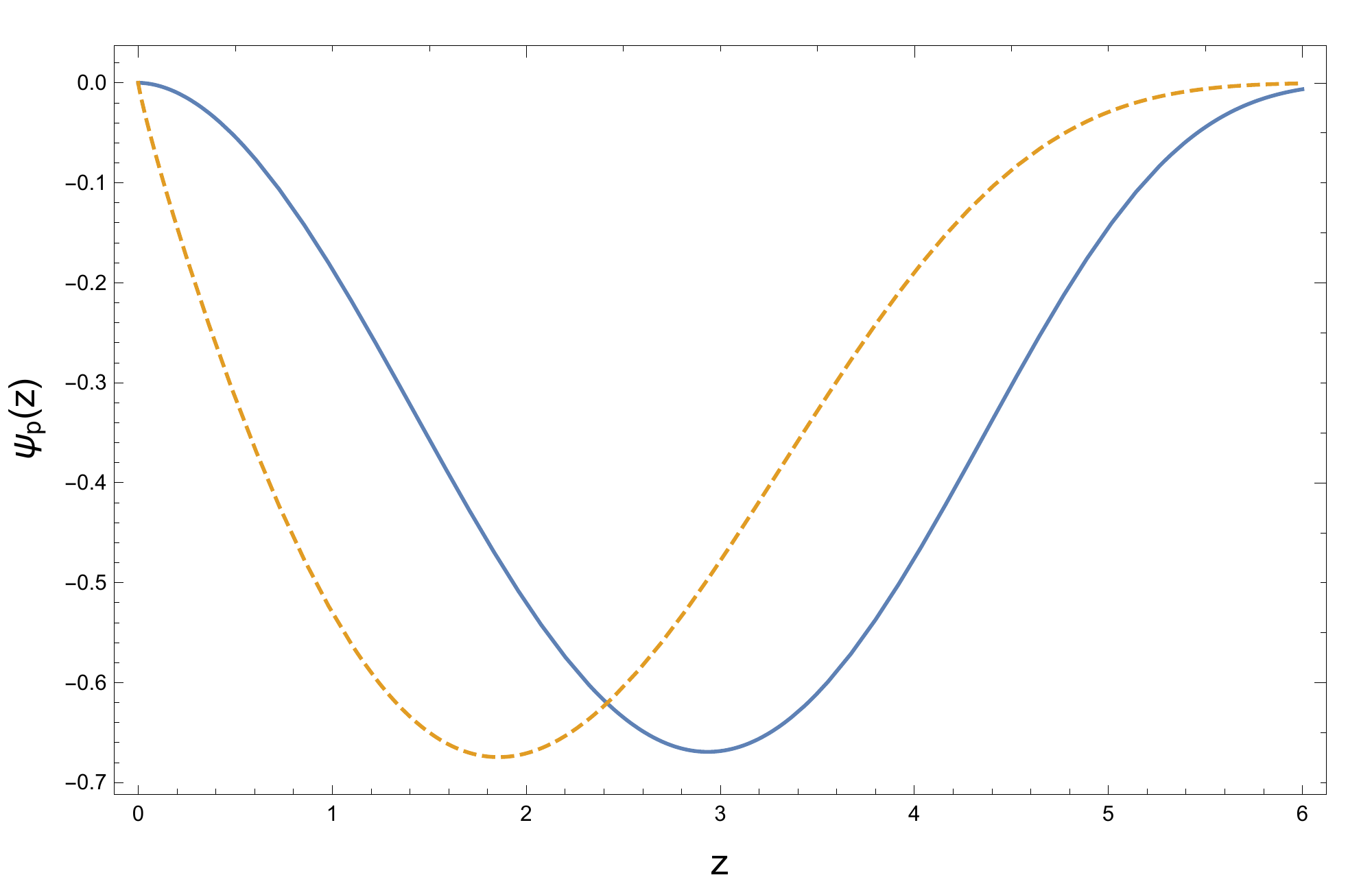} 
	\caption{Chiral wave-functions from Eqs. \eqref{scr} and \eqref{psii} (left with solid line and right with dashed line) for the target proton ($M_p\equiv M_1=0.938$ GeV) using  $k=0.443^2$ GeV$^2$ and $m_5=0.878$ GeV.}
\label{proton}
\end{figure}

 Solving numerically Eq. \eqref{scr} for the ground state ($n=1$) we obtain the target proton wave function $\Psi_i$ shown in Fig. \ref{proton} for both left and right chiralities.

\begin{figure}[!ht]
  \centering
  \includegraphics[scale = 0.70]{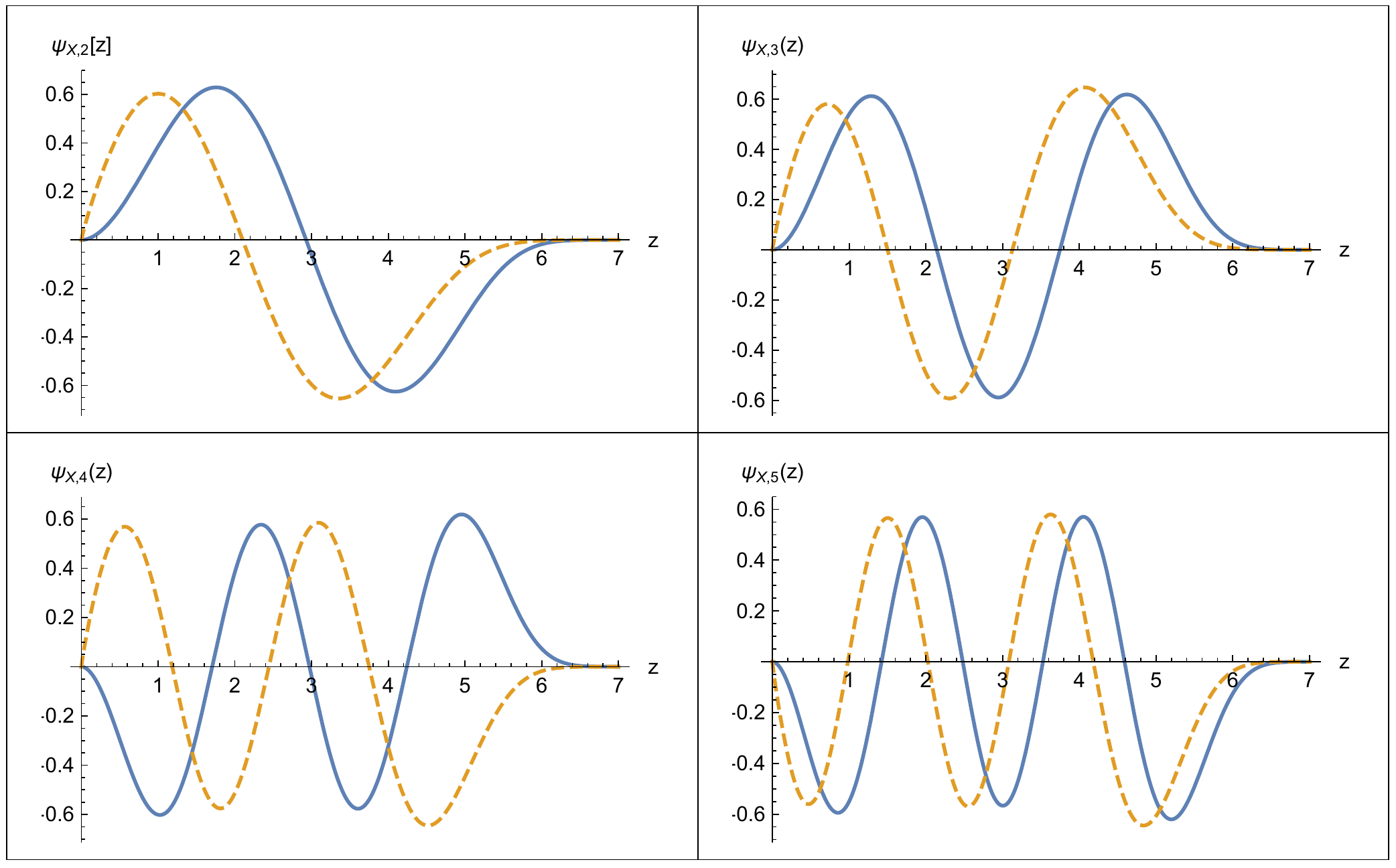} 
\caption{Chiral wave-functions from Eqs. \eqref{scr} and \eqref{psix} for some final excited states with $n=2,3,4,5$,  using $k=0.443^2$ GeV$^2$ and $m_5=0.878$ GeV. In each panel, the left chirality is represented by a solid line and the right chirality by a dashed line. }
\label{HadronX}
\end{figure}

In Fig. \ref{HadronX} we also present the numerical wave functions for both left and right chiralities, obtained by using our model, from Eqs. \eqref{scr} and \eqref{psix}, for some final hadronic states $(n=2,3,4,5)$.

\begin{center}
\begin{figure*}
  \begin{tabular}{c c}
    \includegraphics[width=5. in]{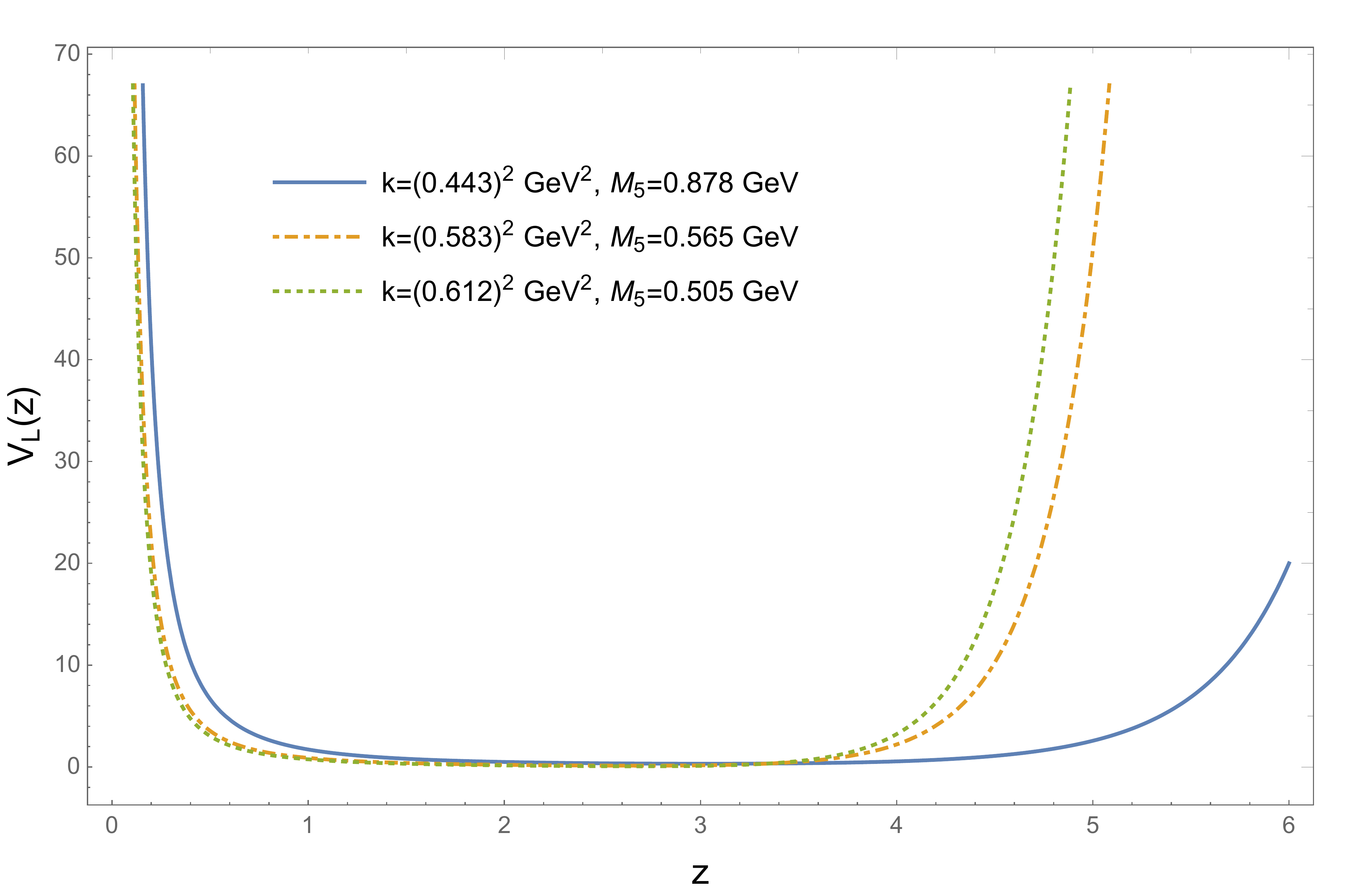}\\
    \includegraphics[width=5. in]{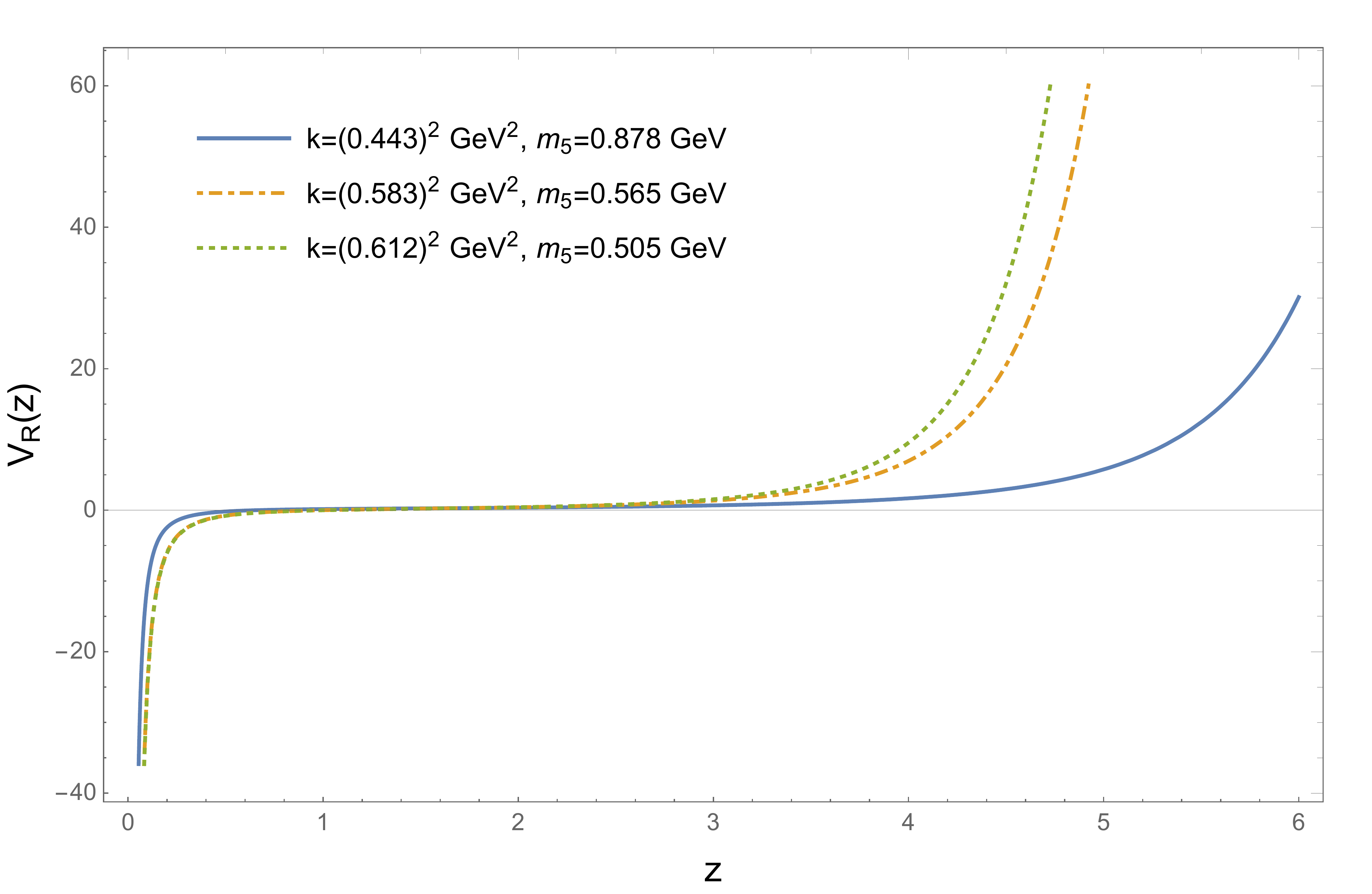} 
  \end{tabular} 
\caption{Chiral potentials given by \eqref{potscr} for the target proton and the final hadronic state X for some values of $k$ and $m_5$.}
\label{potentials}
\end{figure*}
\end{center}

In Fig. \ref{potentials}, we present the potentials considered in the Schr\"odinger equation \eqref{scr} for left and right chiralities, defined by Eq. \eqref{potscr}. The choice of the different values for $m_5$ used in this section will be clarified in section \ref{numerical},  together with our numerical results for the structure functions $F_{1,2}$.  

\section{The DIS interaction action}\label{disintaction}

In this section we will compute explicitly the DIS interaction action. In order to do this, let us recall Eqs. \eqref{corrint} and \eqref{sintdis}, so that:
\begin{eqnarray}\label{funda}
\eta_\mu \langle P + q, s_X | J^{\mu}(q)|P, s\rangle&=& S_{\rm int}\nonumber\\ &=& g_V\,\int{dz\,d^4y\,\sqrt{-g}\,\phi^\mu\,\bar{\Psi}_X\,\Gamma_\mu\,\Psi_i}\,.
\end{eqnarray}
Using the definitions given by Eqs. \eqref{tetra} and \eqref{Cris}, one can write the interaction action $S_{\rm int}$ as:
\begin{eqnarray}\notag
S_{\rm int}&=&g_V\,\int{dz\,d^4y\,\sqrt{-g}\,g^{\mu\,\nu}\phi_\mu\,\bar{\Psi}_X\,e^\alpha_\nu\,\gamma_\alpha\,\Psi_i}\nonumber \\
&=&g_V\,\int{dz\,d^4y\,\sqrt{-g}\,e^{-2A}\,\eta^{\mu\,\nu}\,\phi_\mu\,\bar{\Psi}_X\,e^{A}\,\delta^\alpha_\mu\,\gamma_\alpha\,\Psi_i}\nonumber \\ 
&=&g_V\,\int{dz\,d^4y\,\sqrt{-g}\,\eta^{\mu\nu}\,e^{-A}\,\phi_\mu\,\bar{\Psi}_X\,\gamma_\mu\,\Psi_i}\nonumber \\ \label{int-term}
&=&g_V\,\int{dz\,d^4y\,\frac{e^{2\,k\,z^2}}{z^4}\,\phi^\mu\,\bar{\Psi}_X\,\gamma_\mu\,\Psi_i}.
\end{eqnarray}

\noindent 
The initial and final spinors states, $\Psi_i$ and $\Psi_X$, are given by Eqs. \eqref{psii}
and \eqref{psix}.  
One should note that:
\begin{equation}
 \bar{\Psi}_X=e^{-i\,P_X\cdot\,y}\,z^2\,e^{-k\,z^2}\,\bar{u}_{s_X}(P_X)\left[\left(\frac{1+\gamma_5}{2}\right)\psi^X_L(z)+\left(\frac{1-\gamma_5}{2}\right)\,\psi^X_R(z)\right].   
\end{equation}
With these results and the gauge field $\phi_{\mu}$ given by Eq. \eqref{phimunorm}, using $B\equiv B(z,q)$, we can write the interaction term as follows:
\begin{eqnarray}\notag
S_\text{int}&=&\frac{g_V}{2}\int{d^4y
dz e^{-i\left(P_x-P-q\right)\cdot\,y}\,\eta^\mu \left[\bar{u}_{s_X}\left(\hat{P}_L\,\psi^X_L+\hat{P}_R\,\psi^X_R\right)\gamma_\mu \left(\hat{P}_L\,\psi^i_L+\hat{P}_R\,\psi^i_R\right)\,u_{s_i}\right]B}\\ \notag
&=&\frac{g_V}{2}\,\left(2\,\pi\right)^4\,\delta^4(P_X-P-q)\,\,\eta^\mu\,\int{dz\,\left[\bar{u}_{s_X}\,\gamma_\mu\,\hat{P}_R\,u_{s_i}\psi_L^X\,\psi_L^i\,B+\bar{u}_{s_X}\,\gamma_\mu\,\hat{P}_L\,u_{s_i}\psi_R^X\,\psi_R^i\,B\right]}\\ \label{int-term-2}
&=&\frac{g_V}{2}\,\left(2\,\pi\right)^4\,\delta^4(P_X-P-q)\,\,\eta^\mu\,\left[\bar{u}_{s_X}\,\gamma_\mu\,\hat{P}_R\,u_{s_i}\,\mathcal{I}_L+\bar{u}_{s_X}\,\gamma_\mu\,\hat{P}_L\,u_{s_i}\mathcal{I}_R\right],\label{sintfinal}
\end{eqnarray}

\noindent where the $\mathcal{I}_{R/L}$ are defined in terms of the solutions of the chiral fermions and the solution of the field $B$, so that:
\begin{equation}\
\mathcal{I}_{R/L}=\int{dz\,B(z,q)\,\psi_{R/L}^X(z,P_X)\,\psi_{R/L}^i(z,P)}\,. 
\end{equation}
From Eqs. \eqref{funda} and \eqref{sintfinal}, one gets:
\begin{eqnarray}
\eta_\mu\langle P_X\left|J^\mu(q)\right|P \rangle&=& \left(2\,\pi\right)^4\,\delta^4(P_X-P-q)\,\eta_\mu\langle P+q\left|J^\mu(0)\right|P \rangle\nonumber\\ \label{geff}
&=& \frac{g_{\rm eff}}{2} \delta^4(P_X-P-q)\,\eta_\mu \,\left[\bar{u}_{s_X}\,\gamma^\mu\,\hat{P}_R\,u_{s_i}\,\mathcal{I}_L+\bar{u}_{s_X}\,\gamma^\mu\,\hat{P}_L\,u_{s_i}\mathcal{I}_R\right] 
\\
\eta_\mu\langle P\left|J^\mu(q)\right|P_X \rangle&=&\left(2\,\pi\right)^4\,\delta^4(P_X-P-q)\,\eta_\mu\langle P\left|J^\mu(0)\right|P+q\rangle \nonumber\\
&=& \frac{g_{\rm eff}}{2} \delta^4(P_X-P-q)\,\eta_\nu \,\left[\bar{u}_{s_i}\,\gamma^\mu\,\hat{P}_R\,u_{s_X}\,\mathcal{I}_L+\bar{u}_{s_i}\,\gamma^\mu\,\hat{P}_L\,u_{s_X}\mathcal{I}_R\right].
\end{eqnarray}
\noindent where $g_{\rm eff}$ is an effective coupling constant related to $g_V$. 
Contracting the photon polarization with the hadronic tensor, Eq. \eqref{f4}, one has:
\begin{eqnarray}
\eta_ \mu \eta_ \nu W^{\mu \nu}&=& \frac{\eta_{\mu \nu}}{4} \sum_{M_x^2}\sum_{s_i, s_X} \frac{g^2_{\rm eff}}{4}\,\delta(M^2_X-(P+q)^2) \left[\bar{u}_{s_X}\,\gamma^\mu\,\hat{P}_R\,u_{s_i}\, \bar{u}_{s_i}\,\gamma^ \nu\,\hat{P}_R\,u_{s_X}\,{\cal I}_L^2\right.\nonumber\\ &+&\left. \bar{u}_{s_X}\,\gamma^ \mu\,\hat{P}_R\,u_{s_i}\, \bar{u}_{s_i}\,\gamma^ \nu\,\hat{P}_L\,u_{s_X}\, {\cal I}_L\,{\cal I}_R\, + \,\bar{u}_{s_X}\,\gamma^ \mu\,\hat{P}_L\,u_{s_i}\,\bar{u}_{s_i}\,\gamma^\nu\,\hat{P}_R\,u_{s_X}\,{\cal I}_R\,{\cal I}_L \right. \nonumber\\ &+& \left. \bar{u}_{s_X}\,\gamma^ \mu\,\hat{P}_L\,u_{s_i}\,\bar{u}_{s_i}\,\gamma^\nu\,\hat{P}_L\,u_{s_X}\, {\cal I}_R^2\right]
\end{eqnarray}{}

As we are interested in a spin independent scenario, by using the following property

\begin{equation}
\sum_s  (u_s)_ \alpha (p) \, (\bar u_s)_ \beta (p) = (\gamma^ \mu p_ \mu + M)_{\alpha \beta},
\end{equation}{}

\noindent we perform a summation over the initial and final spin states, $s_i$ and $s_X$, respectively, and then  applying trace engineering, one gets:
\begin{multline}
\eta_ \mu \eta_ \nu W^{\mu \nu}=\frac{g^2_\text{eff}}{4}\,\sum_{M_X^2}\,\delta(M_X^2-(P+q)^2)\,\left\{(\mathcal{I}_L^2+\mathcal{I}_R^2)\left[(P\cdot \eta)^2-\frac{1}{2}\eta\cdot\eta(P^2+P\cdot q)\right]\right. \\
 +\mathcal{I}_L\,\mathcal{I}_R\,M_X^2\,M_0^2\,\eta\cdot\eta\biggr{\rbrace},
\end{multline}

\noindent where we have used $\slashed p = \gamma^ \mu p_ \mu$,  $\{\gamma_5, \gamma_ \mu\} = 0$, and $P_{R/L} \gamma^ \mu = \gamma^ \mu P_{L/R}$.

In order to get the expressions for the structure functions we need to sum over the outgoing states $P_X$, as presented in Eq. \eqref{ImWmunu}. Carrying on this sum to the continuum limit we can evaluate the invariant mass delta function. Following Ref. \cite{Polchinski:2002jw}, this integration will be related to the functional form of the mass spectrum of the produced particles with the excitation number $n$,
\begin{equation*}
    \delta(M_{X}^2-(P+q)^2)\propto \left(\frac{\partial\,M^2_n}{\partial\,n}\right)^{-1}
\end{equation*}
that for the soft and hard wall models  accounts for the lowest state produced at the collision, since the spectrum is linear with $n$  \cite{Polchinski:2002jw, BallonBayona:2007qr}. In our case, this delta will account for $1/ M_X^2$.

Taking into account our choice of transversal polarization ($\eta\cdot q=0$), the hadronic tensor has the following form:
\begin{equation}\label{tensorhadron}
\eta_\mu\,\eta_\nu\,W^{\mu\nu}=\eta^2\,F_1(q^2,x)+\frac{2\,x}{q^2}\,(\eta\cdot P)^2\,F_2^2(q^2,x).    
\end{equation}
From this equation one can construct explicitly the baryonic DIS structure functions, such as:
\begin{eqnarray}\label{F1}
F_1(q^2,x)&=&\frac{{g}^2_\text{eff}}{4}\,\left[M_0\,\sqrt{M_0^2+q^2\left(\frac{1-x}{x}\right)}\,\mathcal{I}_L\,\mathcal{I}_R+\left(\mathcal{I}_L^2+\mathcal{I}_R^2\right)\left(\frac{q^2}{4\,x}+\frac{M_0^2}{2}\right)\right]\frac{1}{M_{X}^2}\\
F_2(q^2,x)&=&\frac{{g}^2_\text{eff}}{8}\frac{q^2}{x}\left(\mathcal{I}_L^2+\mathcal{I}_R^2\right)\frac{1}{M_{X}^2},\label{F2}
\end{eqnarray}
\noindent where $M_{X}\equiv M_{X}(q^2,x)$ is mass of the effective final hadron related to the mass of the initial hadron: 
\begin{equation}
    M_{X}^2(q^2,x) = M_0^2+q^2\,\left(\frac{1-x}{x}\right)\,. 
\end{equation}

Note that the two structure functions are related by: 
\begin{eqnarray}
F_1(q^2,x)
&=&\frac{{g}^2_\text{eff}}{4}\,\frac{M_0}{M_{X}^2}\,\sqrt{M_0^2+q^2\left(\frac{1-x}{x}\right)}\,\mathcal{I}_L\,\mathcal{I}_R 
+ \frac{1}{2} F_2(q^2,x)\left(1+\frac{2xM_0^2}{q^2}\right)\,,
\end{eqnarray}
so that, in the limit of $M_X \gg M_0$, $q \gg M_0$, and $x\to 1$, one finds
\begin{eqnarray}\label{lcg}
F_1(q^2,x)
\approx 
 \frac{1}{2} F_2(q^2,x)\,,
\end{eqnarray}
which behaves like the Callan-Gross relation $2xF_1=F_2$, for $x\to 1$. 

\section{Numerical results for the structure functions}\label{numerical}

In this section we will present our numerical set up and results for the structure functions $F_1(x, q^2)$ and $F_2(x, q^2)$ for some specific values of the Bjorken parameter $x=0.65, 0.75, 0.85$. These values were chosen since they correspond to the highest value of $x$ available experimental data \cite{Whitlow:1991uw, Benvenuti:1989rh}. 

In table \ref{setup}, we present our fit which comes from the numerical solution of Eqs. \eqref{scr} and \eqref{F2}. As discussed before, the bulk mass $m_5$ is a function of $ \Delta_{\rm can}=3/2$ and the anomalous dimension $\gamma$, given by Eq. \eqref{m5ano}. 
 In order to fit the proton mass ($m_p$= 0.938 GeV) and the experimental data for $F_2(x, q^2)$ we have obtained the values for $m_5$,  $k$, $g^2_{eff}$ and $\gamma$ for each value of $x$.

\begin{figure}[!ht]
  \centering
  \includegraphics[scale = 0.35]{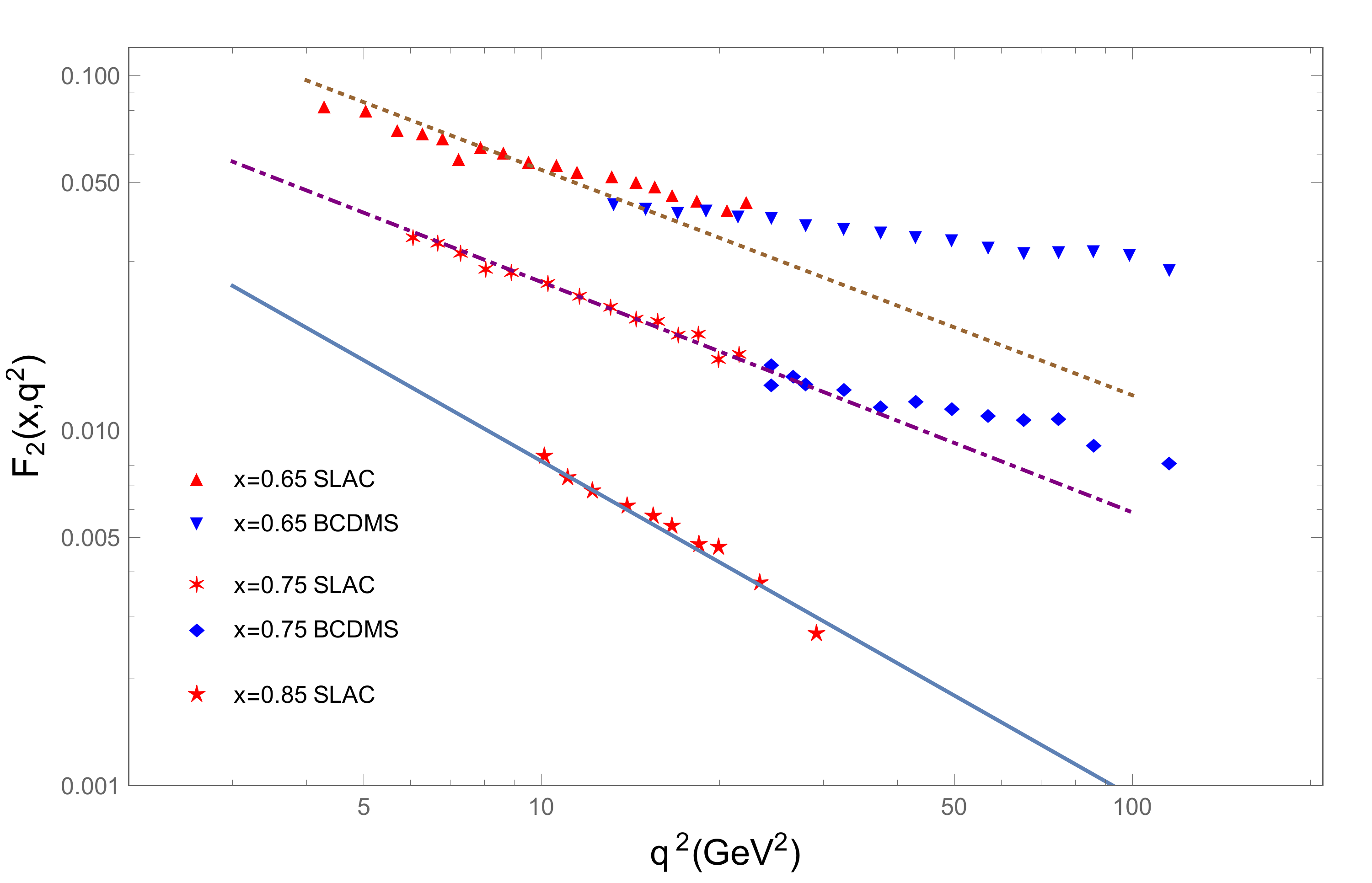} 
\caption{Comparison between experimental data \cite{Whitlow:1991uw, Benvenuti:1989rh} and our results for $F_2(x,q^2)$ as a function of $q^2$ for $x = 0.65$,  $x = 0.75$ and $x = 0.85$ from top to bottom. The dotted, dot-dashed, and solid lines represent our theoretical fits for $x=0.65$, $x=0.75$, and $x=0.85$, respectively. The numerical parameters of the fits are given in Table \ref{setup}.}
\label{results}
\end{figure}

\begin{table}[htb]
\centering
\large
\begin{tabular}{|c||c|c|c|c|}
\hline
$x$ & $m_5$ (GeV)  & $k$ (GeV$^2$) & ${g}^2_{\rm eff}$ & $\gamma$ \\ \hline \hline
0.85 & 0.878 & $0.443^2$ & 1.83 & 0.378  \\ \hline
0.75 & 0.565 &  $0.583^2$ &1.65 & 0.065 \\ \hline
0.65 & 0.505 & $0.612^2$ &3.65 &0.005 \\ \hline
\end{tabular}
\caption{This table summarizes our numerical fit of experimental data. These parameters provide the proton mass as 0.938 GeV and the structure $F_2 (x, q^2)$ shown in Fig. \ref{results}.}
\label{setup}
\end{table}
The Fig. \ref{results} presents our main results. It shows the structure function $F_2(x, q^2)$ against $q^2$ for $x=0.65, 0.75, 0.85$, compared with available experimental data from SLAC \cite{Whitlow:1991uw} and BCDMS \cite{Benvenuti:1989rh} collaborations. One can notice that for $x=0.65, 0.75$ our model deviates from experimental data for very large $q^2$. As expected, our model works better for large $x$.
 
\begin{figure}[!ht]
  \centering
  \includegraphics[scale = 0.35]{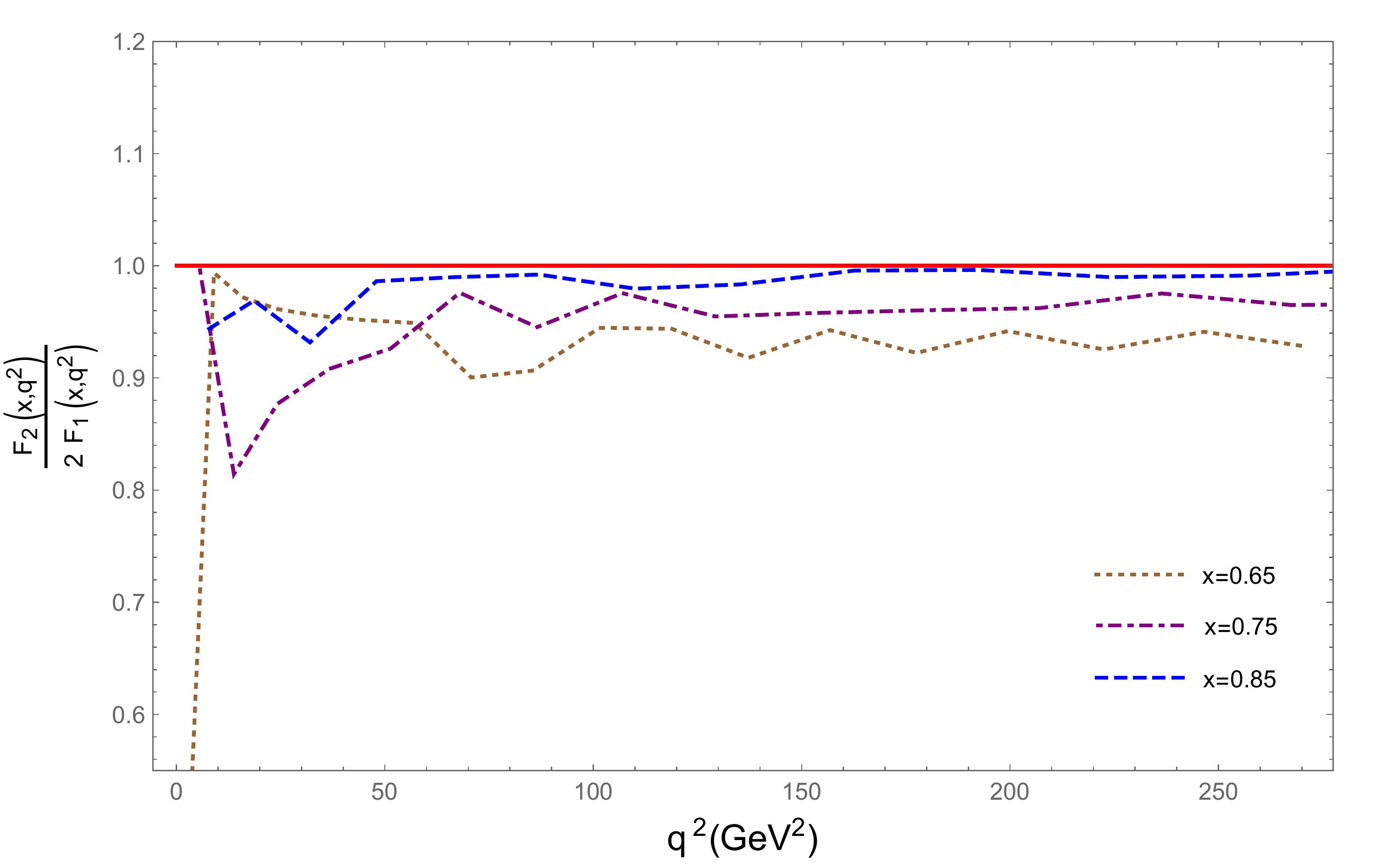}
 \caption{Ratio between $F_2(x, q^2)$ and $2\,F_1(x,q^2)$ from Eqs. \eqref{F1} and \eqref{F2} as a function of $q^2$
 for $x=0.65$, $x=0.75$ and $x=0.85$.  The dotted, dashed, and dot-dashed lines represent out results. The solid line represents the ratio equal to 1.}
\label{cg-085}
\end{figure}

In Fig. \ref{cg-085} we show our results for the ratio $F_2/2\,F_1 $ {\it versus} $q^2$, where one can see that this ratio is approximately equal to one, specially for large $q^2$
 and $x\to 1$, as anticipated by Eq. \eqref{lcg}. It is worth to mention that this is an approximately Callan-Gross relation $F_2=2xF_1$, for $x\to 1$.

\section{Conclusions} \label{conc}

In this section we present our conclusions regarding the results achieved within our holographic description for the baryonic DIS structure functions $F_1(x,q^2)$ and $F_2(x,q^2)$. 
Our AdS/QCD model is characterized by a deformation in AdS space with the introduction of an exponential factor in its metric. One feature of this is that it generates a mass gap for the baryonic sector contrary to the original softwall model. In this approach, the photon has analytical solution while the baryonic fields are numerical. Besides, our model takes into account an anomalous contribution to the canonical scaling dimension of a boundary operator. 

In order to compare with experimental data, we have chosen the target particle as a single proton. Due to the kinematical region and the large $x$ regime we considered the proton to be punctual. This assumption has support on data from PDG showing that the parton distribution functions (PDF) go to zero in the limit of $x\to 1$. 
Our model captures the lepton-proton DIS phenomenology for the range $7 < q^2 < 40$ Gev$^2$, as can be seen in Fig. \ref{results}, for $x=0.65, 0.75, 0.85$. As expected, our model produces better results for large $x$. 

We also found the numerical results for $F_1(x,q^2)$ as can be seen in Fig. \ref{cg-085}, presented through the ratio $F_2/2\,F_1$ as a function of $q^2$. This ratio is similar to the Callan-Gross relation considering $x\to 1$. 

As a final comment, let us mention that the technique developed here for spin $1/2$ baryons could be well extended to baryons with higher spins like $3/2$, $5/2$, etc, despite one does not have experimental data for comparison. 

\begin{acknowledgments}

The authors would like to thanks Alfonso Ballon Bayona for useful discussions. 
E.F.C. and M.A.M.C would like to thank the hospitality of the Jinan University where part of this work was done. E.F.C also would like to thank the hospitality of the Valpara\'iso University where part of this work was done.   A. V. and  M. A. M. C.  would like to thank the financial support given by FONDECYT (Chile) under Grants No. 1180753  and No. 3180592,  respectively. D.L. is supported by the National Natural Science Foundation of China (11805084), the PhD Start-up Fund of Natural Science Foundation of Guangdong Province (2018030310457) and Guangdong Pearl River Talents Plan (2017GC010480). H.B.-F. is partially supported by Coordenação de Aperfeiçoamento de Pessoal de Nível Superior (CAPES),  and Conselho Nacional de Desenvolvimento Científico e Tecnológico (CNPq) under Grant No. 311079/2019-9.

\end{acknowledgments}

\end{document}